\documentstyle[sprocl]{article}

\input{psfig}
\def\Journal#1#2#3#4{{#1} {\bf #2}, #3 (#4)}
\def\preprint#1#2#3{{#1},  {\bf #2},  (#3)}
\def\prep#1#2{{#1} (#2)}

\def\be{\begin{equation}}
\def\ee{\end{equation}}
\def\bea{\begin{eqnarray}}
\def\eea{\end{eqnarray}}

\begin{document}

\title{DO CLUSTER GALAXIES HAVE EXTENDED DARK HALOS?}

\author{Priyamvada Natarajan}

\address{Institute of Astronomy, Madingley Road, Cambridge CB3 0HA,
U. K.}


\maketitle

\abstracts{We present the results of the application of the
methodology introduced by Natarajan \& Kneib (1997) to interpret local
perturbations to the cluster shear field resulting from mass
associated with individual cluster galaxies. The lensing signal is
used to place new constraints on the average mass-to-light ratio and
spatial extents of the dark matter halos associated with
morphologically-classified early-type cluster members in the Abell
cluster AC114. The total mass of a fiducial $L^\ast$ cluster
spheroidal galaxy is found to be largely contained within $\sim$ 15
kpc radius halo ($\sim$~8--10 $R_e$) with a mass-to-light ratio
${M/L_V}\,\sim\,{15^{+10}_{-4}} $ (90 \% c.l.) in solar units within
this radius. Comparisons with similar estimates for field galaxies
suggests that the cluster galaxies in AC114 may possess less extensive
and less massive halos. Additionally, there is some indication that,
at a fixed luminosity, S0 galaxies are less extended than ellipticals,
suggesting a difference in the efficiency of tidal stripping for
different galaxy types. These results enable us to probe the variation
of the mass-to-light ratio with scale via gravitational lensing
methods. Therefore, the prospects for constraining the mass density of
the Universe and understanding galaxy evolution using these
gravitational lensing methods are very promising.}
  
\section{Introduction and Motivation}

The primary motivation for this analysis is to examine the importance
of galaxy-scale halos in defining the distribution of mass in clusters
on intermediate scales ($\simeq 50$ kpc) across a range of
environments within clusters from the core regions to the lower
density outskirts. Using gravitational lensing observations as a
primary tool, our technique considers perturbations associated with an
ensemble of cluster galaxies within a smooth global cluster
potential. The important physical question that one hopes to address
is whether the mass-to-light ratio ($M/L$) of galaxies (measured
within a large effective aperture) varies significantly between high
density cluster regions and the field. Environmental variation of the
$M/L$ ratio of a galaxy might be expected if these galaxies presently
found in dense regions suffered more complex interaction histories
leading to a redistribution of the associated gaseous, stellar and
dark matter components (c.f.\ Moore et al.\ 1996). One possibility is
that the extended dark halo would be preferentially removed and
redistributed, leading to a reduction in the $M/L$ ratio compared to
that found for isolated galaxies of the same morphological
type. However, the scale on which this redistribution occurs (and
hence the `granularity' of the resultant dark matter distribution
within the cluster) is unclear and has important implications for our
understanding of how clusters assemble and evolve.

\section{Brief sketch of the technique}

Our technique involves quantifying the `local' weak lensing induced by
the dark halos around bright cluster galaxies. A measure of these
small-scale perturbations can be derived from the statistical
distribution of the shapes of faint background galaxies. We use the
maximum-likelihood techniques developed in the theoretical discussion
by Natarajan \& Kneib (1995, 1997) and correlate the local
perturbations with cluster galaxies of known morphology and
luminosity. In this way, we provide new constraints on the $M/L$ and
the extent of dark halos in cluster galaxies.  Besides incorporating
both strong and weak lensing constraints simultaneously, further
observational input from the correlations for early-type galaxies that
inhabit the Fundamental Plane are used in the analysis. We have
successfully applied these techniques to a new wide-field {\it Hubble
Space Telescope} ({\it HST}) image of the rich cluster AC114
($z=0.31$). A detailed model of the large-scale mass distribution
within the cluster using the numerous strongly-lensed features visible
in the {\it HST} data is constructed and analyzed.

\begin{figure}
\vspace{-9.0cm}
\centerline{\psfig{file=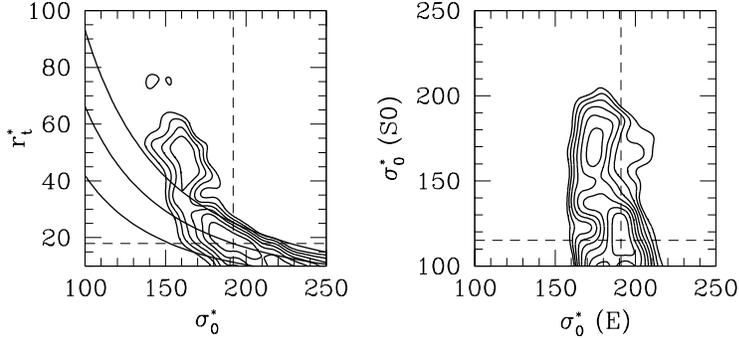,width=10cm}}
\caption{Left Panel: Maximum-likelihood retrieval of the fiducial
parameters for cluster galaxies in AC114: the likelihood peaks at
${r_t^*}\,\sim\,$ 20 kpc and ${\sigma_0^*} \,\sim\,$190 km s${^{-1}}$;
and the inner 3 contour levels correspond respectively to 60\%, 80\%
and 90\% confidence limits The galaxy models used in this case
correspond to constant mass-to-light ratio which are the overplotted
solid curves for $M/L_V$ = 10, 17 \& 23 (increasing from bottom to
top); Right Panel: Parameters recovered for the sub-samples: the
cluster galaxies were split into 2 primary morphological classes the
E's and the S0's. Combined optimization for their respective fiducial
velocity dispersions yields: ${\sigma_0^*}(E)\,\sim\,$ 190 km
s${^{-1}}$ and ${\sigma_0^*}(S0)\,\sim\,$ 120 km s${^{-1}}$. The inner
3 contours correspond to 60\%, 80\% and 90\% confidence limits.}
\end{figure}

\begin{figure}
\vspace{-5.0cm}
\centerline{\psfig{file=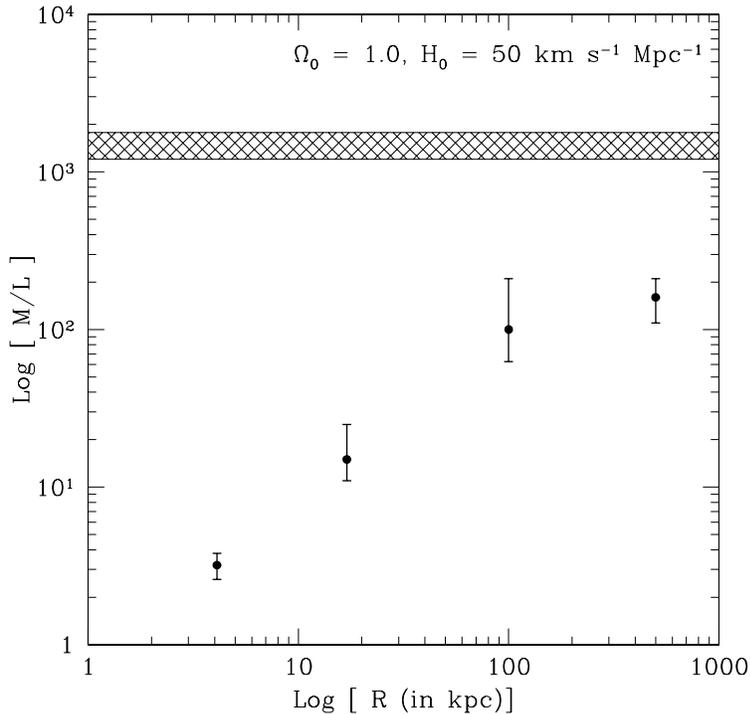,width=10cm}}
\caption{The variation of the M/L ratio with scale as probed by {\bf
gravitational lensing} methods. The point on the smallest scale (4.1
kpc) is from the M/L determined at the Einstein radius for the galaxy
lens HST14176 [Hjorth \& Kneib 1998]; the second point at 17 kpc is
from the galaxy-galaxy lensing in the cluster in AC114 [NKSE 97]; the
third from galaxy-galaxy lensing in the field [BBS 96] and the final
point is the total mass-to-light ratio of the cluster AC114 computed
at 500 kpc [NKSE 97]. The cross-hatched region in the plot is the
global mass-to-light ratio required in order to close the Universe. }
\end{figure}

\section{Results and Discussion}

We report the first detection of the signature of extended dark halos
around galaxies in AC114. ~From our analysis we find that a $L^\ast$
spheroidal cluster galaxy has a total mass of $M \sim
(4.9^{+3.1}_{-1.3}) \times {10^{11}}\,M_\odot$, and a $M/L_V\sim
{15^{+10}_{-4}} (M/L)_\odot$ (90 \% c.l.). We find that the {\bf
total} mass of a fiducial $L^\ast$ cluster spheroidal is primarily
contained with $\sim$ 15 kpc, with some indication that the halos of
the S0 population may be more truncated than those of the
ellipticals. Comparisons with similar estimates for field galaxies
suggests that the cluster galaxies in AC114 may possess less extensive
and less massive halos. Comparing the total mass in cluster galaxies
within 250 kpc of the center (down to the magnitude limit of our
selection criterion) to the total mass of the cluster, we estimate
that approximately 11\% of the mass of the cluster is bound to
individual cluster galaxies. This fraction has important consequences
for the rate of galaxy interactions and hence the evolution of the
cluster on the whole. In Fig.~2, the variation of the mass-to-light
ratio with scale as probed via gravitational lensing using both strong
and weak lensing effects is plotted [also see N. Bahcall's
contribution in this proceeding for comparison with other methods]. A
more detailed discussion of the observed trend in the context of
implications for cosmology and galaxy evolution as well as comparison
with other determinations from purely dynamical methods and X-rays for
estimating the total mass is presented elsewhere (Natarajan, P. 1997).

These first results on the properties of galaxy halos within clusters
from lensing are very encouraging.  We are therefore extending our
analysis using both observations of galaxies in the central regions of
rich clusters at $z=0.17$--0.56 (Natarajan, Kneib \& Smail 1998) and
across a range of environments within a number of clusters at $z\sim
0.3$. These two samples will provide insights into the role of changes
in halo properties in the evolution of both cluster spheroids and disk
galaxies, as well as the variation of these effects with local
environment. The hope of future expansion of this technique appears
good with the proposed installation of the the Advanced Camera for
Survey {\it ACS} in the next {\it HST} Servicing Mission, which will
be able to cover a wider field providing the ideal data-sets for such
studies. Therefore, the prospects for understanding the evolution
of galaxy halos from field galaxies to cluster cores and hence the
radial variation of the mass-to-light ratio are promising.

\section*{Acknowledgments}

Funding from the organizers of the workshop is gratefully
acknowledged. Joachim Wambsganss and Volker Mueller are thanked for
their warm hospitality. Thanks are due to my collaborators - Jean-Paul
Kneib, Ian Smail and Richard Ellis for our many enjoyable and 
useful discussions.

\end{document}